\documentclass[
 ,draft            
  ]
  {aipproc}

\layoutstyle{8x11double}

\begin{document}

\title{Galactic and Extragalactic Magnetic Fields}


\author{Rainer Beck}{address={Max-Planck-Institut f\"ur Radioastronomie,
Auf dem H\"ugel 69, 53121 Bonn, Germany} }

\begin{abstract}
The strength of the total magnetic field in our Milky Way from radio
Zeeman and synchrotron measurements is about $6~\mu$G near the Sun
and several mG in dense clouds, pulsar wind nebulae, and filaments
near the Galactic Center. Diffuse polarized radio emission and
Faraday rotation of the polarized emission from pulsars and
background sources show many small-scale magnetic features, but the
overall field structure in our Galaxy is still under debate. --
Radio synchrotron observations of nearby galaxies reveal dynamically
important magnetic fields of 10--$30~\mu$G total strength in the
spiral arms. Fields with random orientations are concentrated in
spiral arms, while ordered fields (observed in radio polarization)
are strongest in interarm regions and follow the orientation of the
adjacent gas spiral arms. Faraday rotation of the diffuse polarized
radio emission from the disks of spiral galaxies sometimes reveals
large-scale patterns which are signatures of coherent fields
generated by dynamos, but in most galaxies the field structure is
more complicated. -- Strong magnetic fields are also observed in
radio halos around edge-on galaxies, out to large distances from the
plane. The synchrotron scaleheight of radio halos allows to measure
the mean outflow velocity of the cosmic-ray electrons. The ordered
halo fields mostly form an X-shaped pattern, but no large-scale
pattern is seen in the Faraday rotation data. Diffuse polarized
radio emission in the outer disks and halos is an excellent tracer
of galaxy interactions and ram pressure by the intergalactic medium.
-- Intracluster gas can also be significantly magnetized and highly
polarized due to shocks or cluster mergers.
\end{abstract}

\maketitle


\section{1. Introduction}

Most of the visible matter in the Universe is ionized, so that
cosmic magnetic fields are quite easy to generate and, due to the
lack of magnetic monopoles, hard to destroy. Magnetic fields need
illumination by cosmic rays, gas or dust to be detectable. Large
regions in the Universe may be permeated by ``dark'' magnetic
fields. In spite of our increasing knowledge on magnetic fields,
many important questions, especially the origin and evolution of
magnetic fields, their first occurrence in young galaxies, or the
existence of large-scale intergalactic fields remained unanswered.

Results from observations and modeling revealed that magnetic fields
are a major agent in the interstellar and intracluster media. They
affect thermal conduction in galaxy clusters and their evolution
\citep{balbus08}. They contribute significantly to the total
pressure which balances the gas disk against gravitation. They
affect the dynamics of the turbulent interstellar medium (ISM)
\citep{avillez05} and the gas flows in spiral arms \citep{gomez02}.
The shock strength in spiral density waves is decreased and
structure formation is reduced in the presence of strong fields
\citep{dobbs08}. The interstellar fields are closely connected to
gas clouds. Magnetic fields stabilize gas clouds and reduce the
star-formation efficiency to the observed low values
\citep{price08,vazquez05}. On the other hand, magnetic fields are
essential for the onset of star formation as they enable the removal
of angular momentum from the protostellar cloud via ambipolar
diffusion \citep{heitsch04}. {\it MHD turbulence}\ distributes
energy from supernova explosions within the ISM \citep{subra98} and
drives field amplification and ordering via a {\it dynamo}\
\citep{beck96}. {\em Magnetic reconnection}\ is a possible heating
source for the ISM and halo gas \citep{birk98,zimmer97}. Magnetic
fields also control the density and distribution of cosmic rays in
the ISM. Cosmic rays accelerated in supernova remnants can provide
the pressure to drive a {\em galactic outflow}\ and buoyant loops of
magnetic fields via the {\em Parker instability}\ \citep{hanasz02}.
Parker loops can in turn drive a fast dynamo
\citep{hanasz04,parker92}. Outflows from starburst galaxies in the
early Universe may have magnetized the intergalactic medium
\citep{kronberg99}. Understanding the interaction between the gas
and the magnetic field is a key to understand the physics of galaxy
disks and halos and the evolution of galaxies.

\section{2. The tools to measure magnetic fields}

{\em Polarized emission}\ at optical, infrared, submillimeter and
radio wavelengths holds the clue to magnetic fields in galaxies.
Optical polarization is a result of extinction by elongated dust
grains in the line of sight which are aligned in the interstellar
magnetic field (the {\em Davis-Greenstein effect}). The E--vector
points parallel to the field. However, light can also be polarized
by scattering, a process unrelated to magnetic fields and hence a
contamination which is difficult to subtract. Optical polarization
surveys of stars yielded the large-scale structure of the field in
the local and nearby spiral arms of our Galaxy
\citep{wielebinski05}. Polarized diffuse light allowed to map
magnetic field patterns also in nearby galaxies, e.g. in M~51
\citep{scarrott87}.

Polarized emission from elongated dust grains at infrared and
submillimeter wavelengths is not affected by scattered light. The
B--vector points parallel to the magnetic field, useful to map the
structure of magnetic fields in gas clouds \citep{crutcher04} and
the halo of M~82 \citep{greaves00}.

\begin{figure}
\includegraphics[bb = 47 41 522 702,width=0.9\columnwidth,clip=]{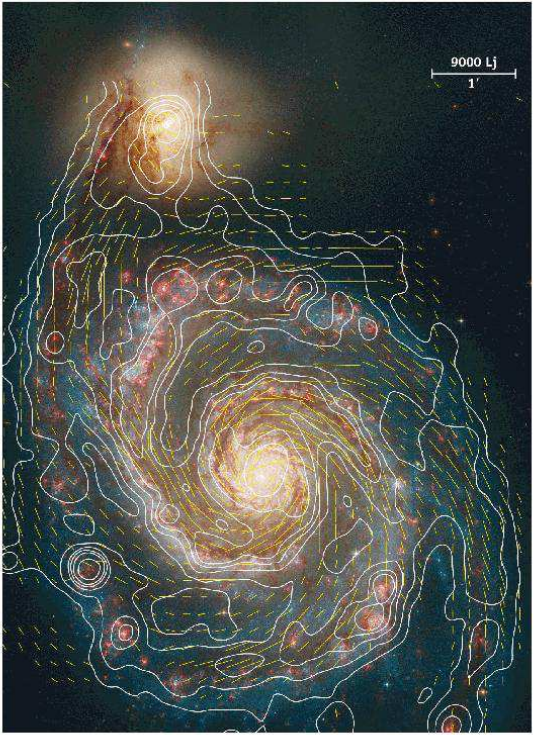}
\caption{HST image of the spiral galaxy M~51, overlaid by contours
of the intensity of total radio emission at 6.2~cm wavelength and
B--vectors, combined from data from the VLA and Effelsberg 100m
telescopes and smoothed to 15'' resolution (Fletcher \& Beck, in
prep.) (Graphics: Sterne und Weltraum. Copyright: MPIfR Bonn and
Hubble Heritage Team).} \label{fig:m51}
\end{figure}

Most of what we know about galactic and intergalactic magnetic
fields comes through the detection of radio waves. {\em Zeeman
splitting}\ of radio spectral lines is the best method to directly
measure the field strength \citep{crutcher99} in gas clouds of the
Milky Way (Sect.~10), OH masers in starburst galaxies
\citep{robishaw08}, and in dense HI clouds in distant galaxies on
the line of sight towards bright quasars \citep{wolfe08}.

The intensity of {\em synchrotron emission}\ is a measure of the
number density of cosmic-ray electrons in the relevant energy range
and of the strength of the total magnetic field component in the sky
plane. Polarized emission emerges from ordered fields. As
polarization ``vectors'' are ambiguous by 180\textdegree{}, they
cannot distinguish {\em regular fields} with a constant direction
within the telescope beam from {\em anisotropic fields}\ which are
generated from turbulent magnetic fields by compressing or shearing
gas flows and frequently reverse their direction on small scales.
Unpolarized synchrotron emission indicates {\em turbulent fields}\
with random directions which have been tangled or generated by
turbulent gas flows.

The intrinsic degree of linear polarization of synchrotron emission
is about 75\%. The observed degree of polarization is smaller due to
the contribution of unpolarized thermal emission, which may dominate
in star-forming regions, by {\em Faraday depolarization}\ along the
line of sight and across the beam \citep{sokoloff98}, and by
geometrical depolarization due to variations of the field
orientation across the beam.

At short radio wavelengths the orientation of the observed B--vector
is parallel to the field orientation, so that the magnetic patterns
of many galaxies could be mapped directly \citep{beck05a}. The
orientation of the polarization vectors is changed in a magnetized
thermal plasma by {\em Faraday rotation}. The rotation angle
increases with the plasma density, the strength of the component of
the field along the line of sight and the square of the observation
wavelength. As the rotation angle is sensitive to the sign of the
field direction, only regular fields can give rise to Faraday
rotation, while anisotropic and random fields do not. For typical
plasma densities and regular field strengths in the interstellar
medium of galaxies, Faraday rotation becomes significant at
wavelengths larger than a few centimeters. Measurements of the
Faraday rotation from multi-wavelength observations allow to
determine the strength and direction of the regular field component
along the line of sight. Its combination with the total intensity
and the polarization vectors can yield the three-dimensional picture
of the magnetic field and allows to distinguish the three field
components: regular, anisotropic and random.

Faraday rotation in foreground objects against the diffuse polarized
background (Milky Way or a radio galaxy) may cause depolarization,
generating a {\em Faraday shadow}\ or {\em Faraday screen}\ which
also allow to estimate the regular field strength
\citep{fomalont89,wolleben04}.

\section{3. Dynamos}

The origin of the first magnetic fields in the Universe is still a
mystery \citep{widrow02}. Protogalaxies probably were already
magnetic due to field ejection from the first stars or from jets
generated by the first black holes \citep{rees05}. A large-scale
primordial field in a young galaxy is hard to maintain because the
galaxy rotates differentially, so that field lines get strongly
wound up during galaxy evolution, in contrast to the observations
which show significant pitch angles. This calls for a mechanism to
sustain and organize the magnetic field. The most promising
mechanism is the dynamo \citep{beck96,brand05} which transfers
mechanical into magnetic energy.

The {\em mean-field $\alpha\Omega$--dynamo}, driven by turbulent gas
motions and differential rotation, generates a large-scale regular
field. Its pattern is described by modes of different azimuthal
symmetry in the disk and vertical symmetry perpendicular to the disk
plane. Several modes can be excited in the same object.

In spherical, rotating bodies like stars, planets or galaxy halos,
the strongest mode is oscillatory and consists of a toroidal field
with a sign reversal across the equatorial plane (vertically {\em
antisymmetric}\ mode A0) and a poloidal dipolar field which is
continuous across the plane but reverses its parity with time.

In flat, rotating objects like galaxy disks, the strongest mode
consists of a toroidal field, which is symmetric with respect to the
plane and has the azimuthal symmetry of an {\em axisymmetric}\
spiral in the plane without sign reversals (vertically {\em
symmetric}\ mode S0) and a weaker poloidal field of quadrupolar
structure with a reversal of the vertical field component across the
equatorial plane \citep{elstner92}. The next strongest mode is of
{\em bisymmetric}\ spiral shape in the plane (mode S1) with two sign
reversals, followed by more complicated modes \citep{bary87}.

The standard mean-field $\alpha\Omega$--dynamo in galaxy disks
amplifies the field and builds up large-scale coherent fields within
a few $10^9$~yr \citep{arshakian08,beck96}. The driving force for
the $\alpha$ effect can be supernova explosions
\citep{ferriere00,gressel08} or cosmic-ray driven Parker loops
\citep{hanasz04,hanasz98,moss99,parker92}.

\begin{figure}
\includegraphics[bb = 34 28 508 588,width=0.9\columnwidth,clip=]{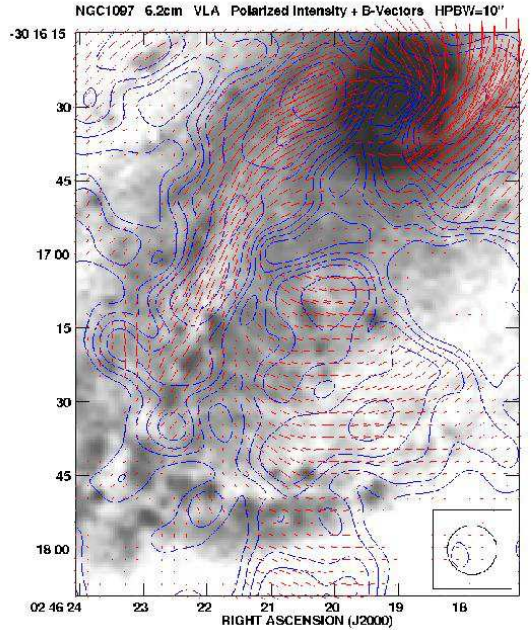}
\caption{Polarized radio emission (contours) and B--vectors of the
barred galaxy NGC~1097, smoothed to 10'' resolution, observed at
6~cm wavelength with the VLA \citep{beck05b}. The background optical
image is from Halton Arp (Copyright: MPIfR Bonn and Cerro Tololo
Observatory).} \label{fig:n1097}
\end{figure}

In a spiral galaxy with disk and halo the more dynamo-active region
determines the global mode, so that either a steady disk mode or an
oscillatory halo mode is expected \citep{moss08}. The oscillation
timescales are very long for galaxies and cannot be determined by
observations.

Kinematical dynamo models including the velocity field of a global
galactic outflow predict field structures which are parallel to the
plane in the inner disk, but open radially outwards, depending on
the outflow speed and direction \citep{brand93}. Some models yield
oscillatory modes, but with a sufficiently fast outflow the
oscillating mode may change into a steady one. However, for very
fast outflows the advection time for the field becomes smaller than
the dynamo amplification time, so that the dynamo action is not
efficient and the field becomes frozen into the flow. Dynamical
models are needed, including the interplay between the gas flow and
the magnetic field.

The mean-field dynamo generates large-scale helicity with a non-zero
mean in each hemisphere. As total helicity is a conserved quantity,
the dynamo is quenched by the small-scale fields with opposite
helicity unless these are removed from the system
\citep{shukurov06}. It seems that {\em outflows}\ are essential for
an effective mean-field dynamo.

The small-scale or {\em fluctuation dynamo}\ \citep{brand05} does
not need general rotation, only turbulent gas motions. It amplifies
weak seed fields to the energy density level of turbulence within
less than $10^9$~yr in galaxy disks \citep{arshakian08,beck94b} and
within about $5~10^9$~yr in the intracluster medium of galaxy
clusters \citep{bertone06,subra06}.

\section{4. Magnetic fields in spiral and barred galaxies}

The strength of the total magnetic field can be determined from the
intensity of the total synchrotron emission, assuming {\em energy
equipartition}\ between the energy densities of the total magnetic
field and the total cosmic rays, with a ratio $K$ between the
numbers of cosmic-ray protons and electrons in the relevant energy
range (usually $K\simeq100$). The equipartition assumption seems to
be valid in galaxies on large scales in space and time, but
deviations occur locally. In regions where electrons lost already a
significant fraction of their energy, e.g. in strong magnetic fields
or radiation fields, or far away from their places of origin, $K$ is
$>100$, so that the standard value of 100 yields an underestimate
\citep{beckkrause05}. On the other hand, in case of fluctuations in
field strength along the line of sight or across the telescope beam,
the equipartition value is an overestimate \citep{beck03}.

The typical average equipartition strength of the total magnetic
field in spiral galaxies is about $10~\mu$G. Radio-faint galaxies
like M~31 and M~33, our Milky Way's neighbors, have weaker total
magnetic fields (about $5~\mu$G), while gas-rich galaxies with high
star-formation rates, like M~51 (Fig.~\ref{fig:m51}), M~83 and
NGC~6946, have average total field strengths of $15~\mu$G. The
strongest fields ($50-100~\mu$G) are found in starburst galaxies,
like M~82 \citep{klein88} and the ``Antennae'' NGC~4038/9
\citep{chyzy04}, and in nuclear starburst regions, like in the
centers of NGC~1097 and other barred galaxies \citep{beck05b}. In
starburst galaxies, however, the equipartition field strength per
average gas surface density is much lower than in normal spirals.
This indicates strong energy losses of the cosmic-ray electrons, so
that the equipartition field strength is probably underestimated by
a factor of a few \citep{thompson06}. Recently, a field strength of
$84~\mu$G was detected in a distant galaxy at z=0.692 by the Zeeman
effect in the HI line seen in absorption against a quasar
\citep{wolfe08}.

The mean energy densities of the magnetic field and of the cosmic
rays in NGC~6946 and M~33 are $\simeq10^{-11}$~erg~cm$^{-3}$ and
$\simeq 10^{-12}$~erg~cm$^{-3}$, respectively
\citep{beck07a,taba08}, about 10 times larger than that of the
ionized gas, but similar to that of the turbulent gas motions across
the whole star-forming disk. The magnetic energy may even dominate
in the outer disk of NGC~6946.

Spiral arms observed in total radio emission appear very similar to
those observed in the far-infrared. The total equipartition field
strength in the arms can be up to $30~\mu$G. The degree of radio
polarization within the spiral arms is only a few \%; hence the
field in the spiral arms must be mostly tangled or randomly oriented
within the telescope beam, which typically corresponds to a few
100~pc. Random fields in spiral arms are probably generated by
turbulent gas motions due to supernovae \citep{avillez05},
turbulence due to spiral shocks \citep{dobbs08} or the turbulent
dynamo \citep{brand05}. In contrast, the ordered (regular and/or
anisotropic) fields traced by the polarized synchrotron emission are
generally strongest ($10-15~\mu$G) in the regions {\em between}\ the
optical spiral arms, oriented parallel to the adjacent optical
spiral arms. In several galaxies the field forms {\em magnetic
arms}\ between the optical arms, like in NGC~6946 \citep{beck07a}.
These are probably generated by the mean-field dynamo (see below).
In galaxies with strong density waves some of the ordered field is
concentrated at the inner edge of the spiral arms, e.g. in M~51
(Fig.~\ref{fig:m51} and \citep{patrikeev06}), but the arm--interarm
contrast of the ordered field is small, much smaller than that of
the random field.

The ordered magnetic field forms spiral patterns in almost every
galaxy \citep{beck05a}, even in ringed galaxies \citep{chyzy08} and
in flocculent galaxies lacking optical spiral structure
\citep{soida02}. Hence, the field lines generally do {\em not}\
follow the (almost circular) gas flow and need dynamo action to
obtain the required radial field components. Spiral fields with
large pitch angles are also observed in central regions of galaxies
and in circum-nuclear gas rings \citep{beck05b}.

On the other hand, in galaxies with massive bars the field lines
seem to follow the gas flow. As the gas rotates faster than the bar
pattern of a galaxy, a shock occurs in the cold gas which has a
small sound speed, while the warm, diffuse gas is only slightly
compressed. As the observed compression of the field in spiral arms
and bars is also small, the ordered field is probably coupled to the
warm diffuse gas and is strong enough to affect its flow
\citep{beck05b}. Here, the polarization pattern is an excellent
tracer of the gas flow in the sky plane and hence complements
spectroscopic measurements. Detailed comparisons between
polarimetric and spectroscopic data are required.

Spiral dynamo modes can be identified from the pattern of
polarization angles and Faraday rotation measures (RM) from
multi-wavelength radio observations of galaxy disks
\citep{elstner92,krause90} or from RM data of polarized background
sources \citep{stepanov08}. The disks of a few spiral galaxies
indeed reveal large-scale RM patterns, as predicted. The Andromeda
galaxy M~31 hosts a dominating axisymmetric disk field (mode S0)
\citep{fletcher04} which extends to at least 25~kpc distance from
the center \citep{han98}. Other candidates for a dominating
axisymmetric disk field are the nearby spiral IC~342
\citep{krause89a} and the irregular Large Magellanic Cloud (LMC)
\citep{gaensler05}. M~81 and PKS~1229-021 are the only candidates
for bisymmetric fields \citep{krause89b,kronberg92}, but the data
quality is limited. The magnetic arms in NGC~6946 can be described
by a superposition of two azimuthal dynamo modes, where the dynamo
wave is phase shifted with respect to the density wave
\citep{beck07a}. However, in many observed galaxy disks no clear
patterns of Faraday rotation were found. Either several dynamo modes
are superimposed and cannot be distinguished with the limited
sensitivity and resolution of present-day telescopes, or the
timescale for the generation of large-scale modes is longer than the
galaxy's lifetime, so that most of the ordered fields traced by the
polarization vectors are anisotropic (with frequent reversals).

Faraday rotation in the direction of QSOs allows to determine the
field pattern in an intervening galaxy
\citep{kronberg92,stepanov08}. This method can be applied to much
larger distances than the analysis of RM of the polarized emission
from the foreground galaxy itself. Faraday rotation of QSO emission
in distant, intervening galaxies revealed significant regular fields
of several $\mu$G strength \citep{bernet08,kronberg08}.

\section{5. The radio--far-infrared correlation}

The correlation between total radio and far-infrared (FIR)
luminosities of star-forming galaxies is one of the tightest
correlations known in astrophysics and a key to understand the role
of magnetic fields in the interstellar medium. The correlation holds
for starburst galaxies \citep{lisenfeld96b} as well as for blue
compact and low-surface brightness galaxies \citep{chyzy07}. It
extends over five orders of magnitude \citep{bell03} and is valid to
redshifts of at least 3 \citep{seymour08}. Only galaxies with very
recent starbursts reveal significantly smaller radio-to-FIR ratios
\citep{roussel03}. Slightly different slopes of the correlation are
obtained if the FIR emission is separated into that from warm and
cold dust \citep{pierini03}.

\begin{figure}
\includegraphics[bb = 27 25 516 493,width=0.9\columnwidth,clip=]{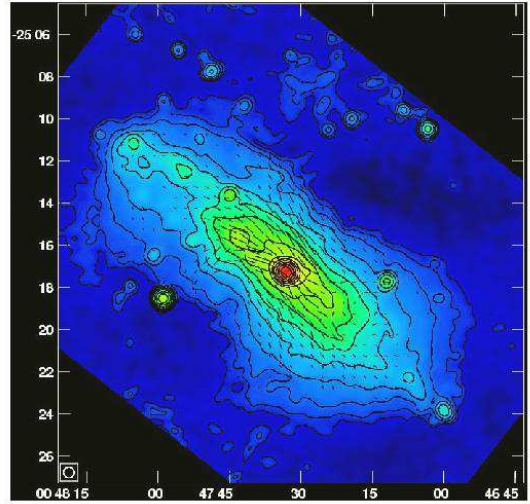}
\caption{Total radio emission at 6~cm wavelength (30'' resolution)
and B--vectors of the almost edge-on spiral galaxy NGC~253, combined
from observations with the VLA and the Effelsberg 100m telescope
\citep{heesen08,heesen05} (Copyright: AIRUB Bochum and MPIfR Bonn) }
\label{fig:n253}.
\end{figure}

Strongest total radio emission (tracing the total, mostly turbulent
field) generally coincides with highest emission from dust and gas
in the spiral arms: The correlation also holds for the local radio
and FIR or mid-IR (MIR) intensities within galaxies
\citep{hoernes98,patrikeev06,taba07,vogler05,walsh02}. In NGC~6946,
the highest correlation of all spectral ranges is between the total
radio intensity at $\lambda$6~cm and the mid-infrared dust emission,
while the correlation with the cold gas (as traced by the CO(1-0)
transition) is less tight
\citep{frick01b,nieten06,vogler05,walsh02}. A wavelet
cross-correlation analysis for M~33 showed that the radio--FIR
correlation holds for all scales down to 1~kpc \citep{taba07}. The
correlation breaks down below scales of about 50~pc \citep{hughes06}
and in radio halos (Fig.~\ref{fig:n253}). Note that the polarized
intensity (tracing the ordered field) is anticorrelated or not
correlated with all tracers of star formation \citep{frick01b}.

It is not obvious why the dominating nonthermal radio and the
thermal far-infrared intensities are so closely related. The
intensity of synchrotron emission depends not only on the density of
cosmic-ray electrons (CRE) which are accelerated in supernova
remnants, but also on about the square of the strength of the total
magnetic field $B_\mathrm{t}$ (its component in the sky plane, to be
precise). The radio--FIR correlation requires that magnetic fields
and star-formation processes are connected. If $B_\mathrm{t}$ is
strong, most of the cosmic-ray energy is released via synchrotron
emission within the galaxy, the CRE density decreases with
$B_\mathrm{t}^2$ and the integrated radio synchrotron luminosity
depends on the CRE injection rate, not on $B_\mathrm{t}$. If the
thermal energy from star formation is also emitted within a galaxy
via far-infrared emission by warm dust, this galaxy can be treated
as a ``calorimeter'' for thermal and nonthermal emission. Prime
candidates for ``calorimeter'' galaxies are those with a high
star-formation rate (SFR). If $B_\mathrm{t}$ increases with SFR
according to $B_\mathrm{t}\propto SFR^{0.5}$, a linear radio--FIR
correlation for the integrated luminosities is obtained
\citep{lisenfeld96a,lisenfeld96b,voelk89}. However, the calorimeter
model cannot explain the local correlation within galaxies.

In galaxies with low or medium SFR, synchrotron energy losses are
probably unimportant and cosmic-ray electrons can leave the galaxy.
To obtain a global or local radio--FIR correlation, coupling of
magnetic fields to the gas clouds is needed. A scaling
$B_\mathrm{t}\propto\rho^{1/2}$ was proposed
\citep{helou93,hoernes98,niklas97} where $\rho$ is the average
density of the neutral gas (atomic + molecular), given by the
average number density of clouds within the telescope beam (not to
be confused with the scaling for the internal density of molecular
clouds as derived from Zeeman measurements, see
Fig.~\ref{fig:zeeman}). A nonlinear correlation (with a slope of
about 1.3) between the nonthermal radio luminosity and the
far-infrared luminosity from warm dust is achieved by further
assuming energy equipartition between magnetic fields and cosmic
rays and a Schmidt law of star formation ($SFR\propto\rho^{1.5}$)
\citep{niklas97}. In this model the total magnetic field strength
and the star-formation rate $SFR$ are related via
$B_\mathrm{t}\propto SFR^{1/3}$.

\section{6. Radio halos of edge-on galaxies}

\begin{figure}
\includegraphics[bb = 26 27 457 609,width=0.9\columnwidth,clip=]{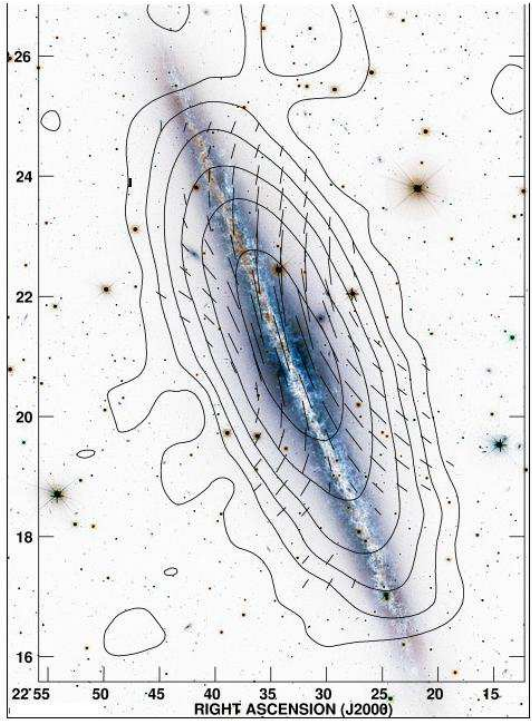}
\caption{Total radio emission at 3.6~cm wavelength (84'' resolution)
and B--vectors of the edge-on spiral galaxy NGC~891, observed with
the Effelsberg 100m telescope \citep{krause08}. The background
optical image is from the CFHT (Copyright: MPIfR Bonn and
CFHT/Coelum).} \label{fig:n891}
\end{figure}

Radio halos are observed around the disks of many edge-on galaxies
\citep{dumke95,hummel91b,irwin99}, but their radio intensity and
extent varies significantly. The halo luminosity in the radio range
correlates with those in H$\alpha$ and X-rays \citep{tuell06},
although the detailed halo shapes vary strongly between the
different spectral ranges. These results suggest that star formation
in the disk is the energy source for halo formation and the halo
size is determined by the energy input from supernova explosions per
surface area in the projected disk \citep{dahlem95}.

\begin{figure}
\includegraphics[bb= 26 28 465 380,width=0.95\columnwidth,clip=]{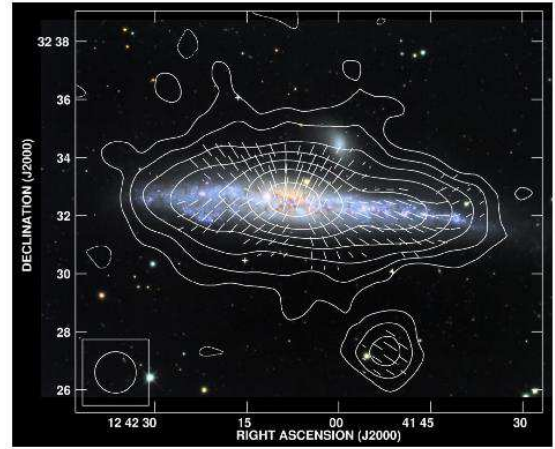}
\caption{Total radio emission at 3.6~cm wavelength (84'' resolution)
and B--vectors of the edge-on spiral galaxy NGC4631, observed with
the Effelsberg 100m telescope \citep{krause07}. The background
optical image is from the Misti Mountain Observatory (Copyright:
MPIfR Bonn).} \label{fig:n4631}
\end{figure}

In spite of their different intensities and extents, the scale
heights of radio halos at 5~GHz are $\simeq1.8$~kpc
\citep{dumke98,heesen08} with a surprisingly small scatter. Their
sample of galaxies included one of the weakest halos, NGC~4565, as
well as the brightest ones known, NGC~891 (Fig.~\ref{fig:n891}) and
NGC~253 (Fig.~\ref{fig:n253}). In case of energy equipartition, the
scale height of the total field is at least $(3+\alpha)$ times
larger than the synchrotron scale height (where $\alpha\simeq1$ is
the synchrotron spectral index), hence $\ge7$~kpc on average. A
prominent exception is NGC~4631 with the largest radio halo (field
scale height of $\ge10$~kpc) observed so far
\citep{hummel91a,hummel90,krause04} (Fig.~\ref{fig:n4631}). Due to
the large scale heights, the magnetic energy density in halos is
much higher than that of the thermal gas \citep{ehle98}, while still
lower than the dominating kinetic energy of the wind outflow.

The magnetic field scale heights are lower limits because the
cosmic-ray electrons lose their energy with height above the plane,
so that the equipartition formula yields too small values for the
field strength \citep{beckkrause05}. The scale height of the ordered
field may be even larger because the degree of linear polarization
increases with height above the plane \citep{dumke98}.

Radio halos grow in size with decreasing observation frequency which
indicates that the extent is limited by energy losses of the
cosmic-ray electrons, i.e. synchrotron, inverse Compton and
adiabatic losses \citep{pohl90}. The stronger magnetic field in the
central regions leads to larger synchrotron loss, leading to the
``dumbbell'' shape of many halos, e.g. NGC~253
\citep{beck94a,heesen05} (Fig.~\ref{fig:n253}), which is in contrast
to its almost spherical X-ray halo \citep{pietsch00}. From the radio
scale heights at three frequencies and the corresponding electron
lifetimes (due to synchrotron, IC and adiabatic losses) a transport
speed of about 300~km/s was measured in the halo of NGC~253
\citep{heesen08}. The similar scale height of the radio halos around
most edge-on galaxies observed so far, in spite of the different
field strengths and hence different electron lifetimes, indicates
that the outflow speed increases with the average field strength and
the star-formation rate.

The exceptionally large radio halo around the irregular and
interacting galaxies M~82 \citep{reuter92} and NGC~4631
\citep{hummel90} with their mostly radial fields in the inner region
\citep{krause08,reuter94} (Fig.~\ref{fig:n4631}) indicates that the
wind transport is more efficient here than in other galaxies. The
gravitational potential or external forces may also play a role to
determine the outflow velocity.

\section{7. Magnetic field structure and Faraday rotation in radio halos}

Radio polarization observations of nearby galaxies seen edge-on
generally show a disk-parallel field near the disk plane
\citep{dumke95}. High-sensitivity observations of several edge-on
galaxies like NGC~891 \citep{krause07} (Fig.~\ref{fig:n891}),
NGC~5775 \citep{tuell00}, NGC~253 \citep{heesen08}
(Fig.~\ref{fig:n253}) and M~104 \citep{krause06} show vertical field
components which increase with increasing height $z$ above and below
the galactic plane and also with increasing radius \citep{krause07}.
These so-called X-shaped magnetic fields contain strong radial
components.

The observation of X-shaped field patterns is of fundamental
importance to understand the field origin in halos. The X-shape is
inconsistent with the predictions from standard dynamo models
without outflows (see Sect.~3). The field is probably transported
from the disk into the halo by an outflow emerging from the disk. A
superwind emerging from a starburst exists only in one of the
edge-on galaxies observed so far, NCC~253 \citep{heckman90}, and
should produce a bipolar field pattern in the inner halo which is
however not observed. A recent model for global outflows from galaxy
disks (neglecting magnetic fields) shows a X-shaped velocity field
\citep{vecchia08} which may drag out the magnetic field. Improved
models including magnetic fields and dynamo action are needed.

The detailed analysis of the highly inclined galaxy NGC~253
(Fig.~\ref{fig:n253}) allowed a separation of the observed field
into an axisymmetric disk field and a halo field inclined by about
50\textdegree{} \citep{heesen08}. Similar tilt angles are also
observed at large heights in other edge-on galaxies.

In the radio halos of M~82 \citep{reuter92} and NGC~4631
\citep{golla94} (Fig.~\ref{fig:n4631}) a few magnetic spurs could be
resolved, connected to star-forming regions. These observations
support the idea of a strong galactic outflow which is driven by
regions of star formation in the inner disk.

Polarization ``vectors'' do not distinguish between a halo field
which is sheared into elongated Parker loops or a regular field. A
large-scale regular field can be measured only by Faraday rotation
measures (RM). RM values in halos, e.g. in NGC~253 \citep{beck94a}
and in NGC~4631 \citep{krause04}, do not show large-scale patterns,
as predicted from dynamo models.

Faraday depolarization is another method to detect magnetic fields
and ionized gas in galaxy halos. In NGC~891 and NGC~4631 the mean
degree of polarization at 1.4~GHz increases from about 1\% in the
plane to about 20\% in the upper halo \citep{hummel91a}. This was
modeled by depolarization due to random magnetic fields of $10~\mu$G
and $7~\mu$G strength, ionized gas with scale heights of 0.9~kpc and
1.3~kpc, and densities in the plane of 0.03~cm$^{-3}$ and
0.07~cm$^{-3}$, respectively.

\section{8. Interactions}

Interaction between galaxies or with the intergalactic medium
imprints unique signatures onto magnetic fields in disks and halos.
The Virgo cluster is a location of strong interaction effects.
Highly asymmetric distributions of the polarized emission shows that
the magnetic fields of several spirals are strongly compressed on
one side of the galaxy \citep{vollmer07,wez07}.

\begin{figure}
\includegraphics[width=0.9\columnwidth]{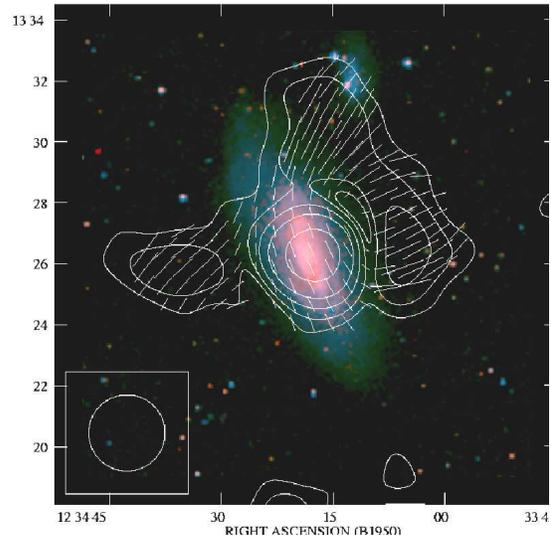}
\caption{Polarized radio emission (2.5' resolution) and B--vectors
of the spiral galaxy NGC~4569 in the Virgo Cluster, observed at 6~cm
wavelength with the Effelsberg 100m telescope \citep{chyzy06}
(Copyright: Cracow Observatory).} \label{fig:n4569}
\end{figure}

Interaction may also induce violent star-formation activity in the
nuclear region or in the disk which may produce huge radio lobes due
to outflowing gas and magnetic field. The lobes of the Virgo spiral
NGC~4569 reach out to at least 25~kpc from the disk and are highly
polarized \citep{chyzy06} (Fig.~\ref{fig:n4569}). However, there is
no indication for a recent starburst, so that the radio lobes are
probably a signature of activity in the past.

Hence, polarized radio emission is an excellent tracer of
interactions. As the decompression timescale of the field is very
long, it keeps memory of events in the past. These are still
observable if the lifetime of the illuminating cosmic-ray electrons
is sufficiently large. Radio observations at low frequencies are
preferable.

\section{9. Galaxy clusters and the intergalactic medium}

\begin{figure}
\includegraphics[bb = 127 164 478 588,width=0.9\columnwidth,clip=]{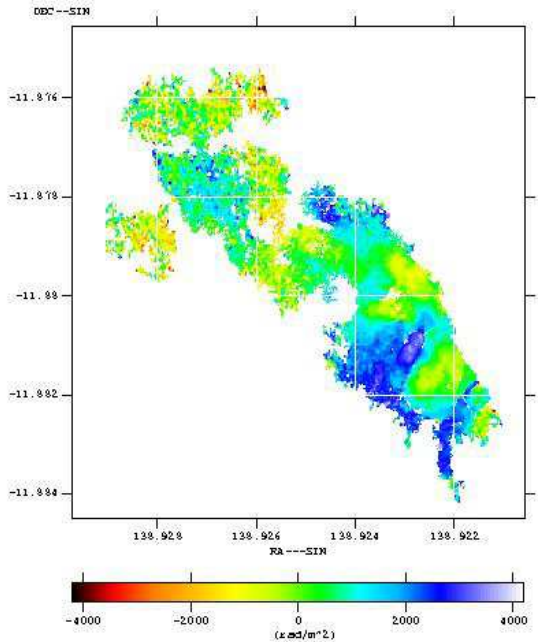}
\caption{Faraday rotation map of the northern lobe of the radio
galaxy Hydra~A embedded in a cluster \citep{taylor93} (Copyright:
NRAO).} \label{fig:hydra}
\end{figure}

Some fraction of galaxy clusters, mostly the X-ray bright ones, has
diffuse radio emission \citep{cassano08}. Radio {\em halos}\ are
mostly unpolarized and emerge from turbulent intracluster magnetic
fields while {\em relics}\ can be highly polarized due to merger
shocks \citep{ensslin98,giovannini91,govoni04}. Magnetic fields
ordered at about 1~Mpc scale were discovered in Abell~2255
\citep{govoni05}. Equipartition strengths of the total magnetic
field range from 0.1 to $1~\mu$G in halos and are higher in relics.
On the other hand, Faraday rotation (RM) data indicate much
stronger, regular fields in clusters of up to $8~\mu$G strength
\citep{govoni04}, even $40~\mu$G in the cores of cooling flow
clusters \citep{carilli02} and may be dynamically important. The
reason for the discrepancy in field strengths is still under
discussion \citep{beckkrause05,carilli02,govoni04}. The origin of
cluster fields could be outflows from AGNs, turbulent wakes or
cluster mergers, possibly amplified by a turbulent dynamo
\citep{bertone06,subra06}. High-resolution RM maps of radio galaxies
within clusters (Fig.~\ref{fig:hydra}) allow to derive the
turbulence spectra of the intracluster magnetic fields which are of
Kolmogorov type \citep{vogt03,vogt05}.

The search for magnetic fields in the intergalactic medium (IGM) is
of fundamental importance for cosmology. All ``empty'' space in the
Universe may be magnetized. Its role as the likely seed field for
galaxies and clusters and its possible relation to structure
formation in the early Universe, places considerable importance on
its discovery. Models of structure formation predict strong
intergalactic shocks which enhance the field. Fields of
$B\simeq10^{-9}$--$10^{-8}$~G are expected along filaments of 10~Mpc
length with $n_e\simeq10^{-5}$~cm$^{-3}$ electron density
\citep{kronberg06} which yield weak synchrotron emission and Faraday
rotation. Their detection is a big challenge and should become
possible with LOFAR and the SKA (Sect.~12).

To date there has been no detection of a general magnetic field in
the IGM. Current upper limits on the average strength of the regular
IGM field from Faraday rotation data suggest $|B_{\rm IGM}| \le
10^{-8}-10^{-9}$~G \citep{kronberg94}. In an intergalactic region of
about 2\textdegree{} extent west of the Coma Cluster, containing a
group of radio galaxies, enhanced synchrotron emission yields an
equipartition total field strength of 0.2-0.4~$\mu$G
\citep{kronberg07}. Such fields may be typical for intergalactic
filaments and should be observable with LOFAR and SKA (Sect.~12).

\section{10. Magnetic field strength in the Milky Way}

\begin{figure}
\includegraphics[width=0.9\columnwidth]{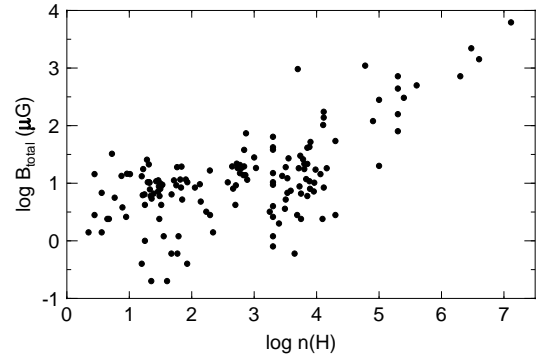}
\caption{Compilation of present-day Zeeman measurements of the
magnetic field in gas clouds plotted against the hydrogen volume
density $n(H)$ (in cm$^{-3}$). To derive the total field
$B_{total}$, each measured line-of-sight component was multiplied by
a factor of 2 which is the average correction factor for a large
sample (from \citep{crutcher07}) (Copyright: R.~M.~Crutcher).}
\label{fig:zeeman}
\end{figure}

Zeeman measurements of HI and molecular lines in gas clouds reveal
the line-of-sight component of the cloud's magnetic field. A
compilation of existing data is shown in Fig.~\ref{fig:zeeman}. The
average strength for cloud densities $n$ below about $10^3$ is about
$6.0~\pm1.8~\mu$G \citep{heiles05}, consistent with results from
synchrotron emission (see below). The energy density of the magnetic
field is similar to that of the turbulent clouds motions, but larger
than the thermal energy density \citep{heiles05}, similar to the
results for external galaxies (Sect.~4). For larger densities, the
field strength scales with $n^{0.65\pm0.05}$ \citep{crutcher07}.
Zeeman splitting of OH maser lines from dense clouds yield field
strengths of a few mG \citep{fish03}. In dense dust clouds field
strengths of about $100~\mu$G were measured from submillimeter
polarimetry \citep{crutcher04}.

\begin{figure}
\includegraphics[width=0.8\columnwidth]{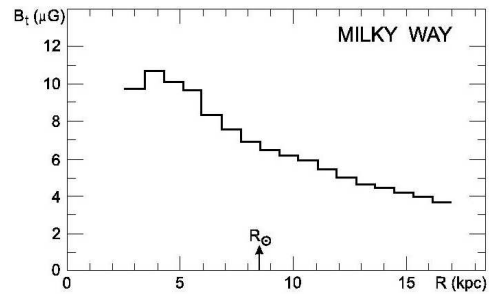}
\caption{Equipartition magnetic field strength, derived from radio
synchrotron emission (from Berkhuijsen, in \citep{wielebinski05})
(Copyright: MPIfR Bonn).} \label{fig:bfield}
\end{figure}

Radio synchrotron data also yield an equipartition strength of the
total field of 6~$\mu$G near the Sun and about 10~$\mu$G in the
inner Galaxy (Fig.~\ref{fig:bfield}). In our Galaxy the accuracy of
the equipartition assumption can be tested, because we have
independent information about the local cosmic-ray energy density
from in-situ measurements and about their radial distribution from
$\gamma$-ray data. Combination with the radio synchrotron data
yields a local strength of the total field of 6~$\mu$G
\citep{strong00}, the same value as derived from energy
equipartition. The radial scale length of the equipartition field of
$\simeq12$~kpc is also similar to that in \citep{strong00}. In the
nonthermal filaments near the Galactic center the field strength may
reach several 100~$\mu$G \cite{reich94,yusef96}, but the pervasive
diffuse field is much weaker \citep{novak05}, probably 20--40~$\mu$G
\citep{larosa05} (scaled for a proton/electron ratio of 100).
Milligauss fields were also determined in the pulsar wind nebula
DA~495 from a break in its synchrotron spectrum \citep{kothes08}.

Synchrotron polarization observations in the local Galaxy imply a
ratio of ordered to total field strengths of $\simeq0.6$
\citep{berk71,heiles96}. For a total field of 6~$\mu$G these results
give 4~$\mu$G for the local ordered field component (including
anisotropic random fields).

Rotation measure (RM) and dispersion measure data of pulsars give an
average strength of the local coherent regular field of
$1.4\pm0.2~\mu$G \citep{han94,rand94}. In the inner Norma arm, the
average strength of the coherent regular field is $4.4\pm0.9~\mu$G
\citep{han02}. The regular field is claimed to be stronger in the
Galactic spiral arms than in the interarm regions \citep{han06},
this is however in contrast to the results from external galaxies
(Sect.~4).

The values for the regular field strength derived from pulsar data
are smaller than the equipartition estimates derived from polarized
intensities. The latter overestimate the strength of the coherent
regular field if anisotropic turbulent fields (Sect.~2) are present,
but the former can also be overestimates if the small-scale
fluctuations in field strength and in electron density are
correlated \citep{beck03} which is indicated by the analysis of RM
data from extragalactic sources \citep{sun08}.

\section{11. Magnetic field structure of the Milky Way}

\begin{figure}
\includegraphics[bb = 55 49 522 373,width=0.9\columnwidth,clip=]{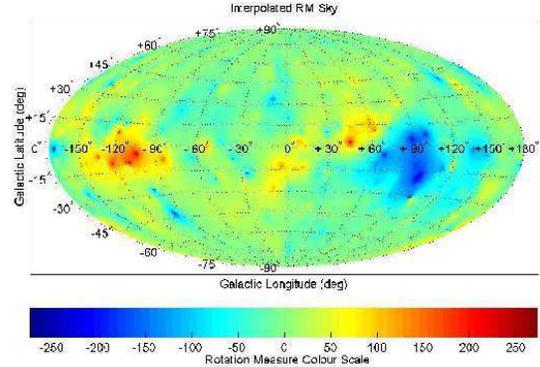}
\caption{All-sky map of rotation measures in the Milky Way,
generated from rotation measures towards about 800 polarized
extragalactic sources, smoothed to a resolution of one square degree
(from \citep{johnston04}) (Copyright: M.~Johnston-Hollitt).}
\label{fig:johnston}
\end{figure}

\begin{figure}
\includegraphics[bb = 26 28 378 380,width=0.95\columnwidth,clip=]{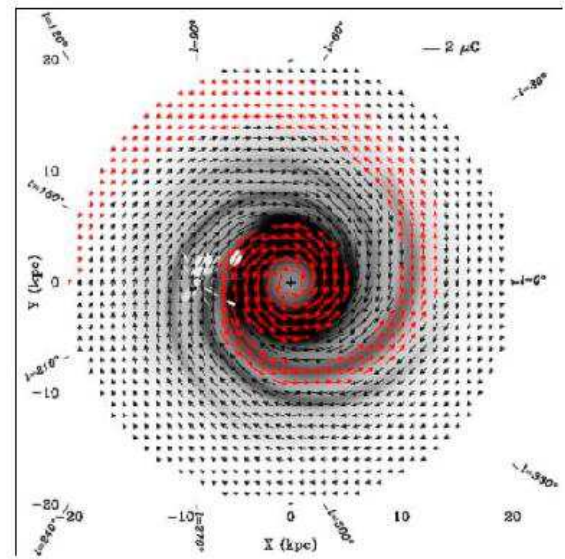}
\caption{The axisymmetric model of the large-scale structure of
magnetic fields in the Milky Way disk, derived from polarization
surveys and rotation measures of extragalactic sources (from
\citep{sun08}) (Copyright: MPIfR Bonn).} \label{fig:sun}
\end{figure}

The all-sky distribution of extragalactic rotation measures (RM)
(Fig.~\ref{fig:johnston}) shows the component of the large-scale
field along the line of sight. The main RM peaks near the Galactic
plane at longitudes of +100\textdegree{} and -110\textdegree{} show
that the local Galactic field is oriented mainly parallel to the
plane.

The maps of polarized synchrotron emission from the Milky Way from
DRAO, Villa Elisa and WMAP and the new Effelsberg RM survey of
polarized extragalactic sources were used to model the large-scale
Galactic field \citep{sun08} (Fig.~\ref{fig:sun}). One large-scale
reversal is required about 1--2~kpc inside the solar radius, which
agrees with the detailed study of 148 RMs from extragalactic sources
near the southern Galactic plane \citep{brown07}.

Whether the sign distribution of Zeeman measurements supports a
large-scale regular field in the Galaxy is still under discussion
\citep{fish03,han07}.

Rotation measure from pulsars allow a more detailed investigation of
the field structure in the Milky Way. The field reversal inside the
solar radius is also seen in the pulsar data
\citep{frick01a,han02,han06,rand94}. The local field runs clockwise
(seen from the northern Galactic pole) but reverses to
counter-clockwise towards the next inner spiral arm, the
Sagittarius-Carina arm, and is still counter-clockwise in the inner
molecular ring.

\begin{figure}
\includegraphics[bb = 34 28 342 244,width=0.95\columnwidth,clip=]{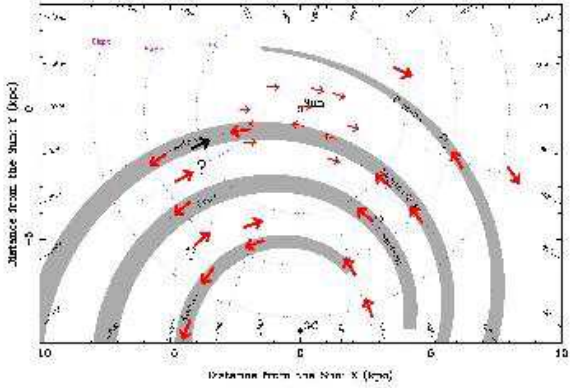}
\caption{The bisymmetric model of the large-scale structure of
magnetic fields in the Milky Way disk, derived from pulsar rotation
measures (from \citep{han06}) (Copyright: J.~L.~Han).}
\label{fig:han}
\end{figure}

Fig.~\ref{fig:han} shows the location of pulsars with measured RMs
within the Galactic plane and the derived directions of the
large-scale field. The Sun is located between two spiral arms, the
Sagittarius/Carina arm and the Perseus arm. The mean pitch angle of
the arms is about -18\textdegree{} for the stars and
-13\textdegree{} for all gas components \citep{vallee02}. Starlight
polarization and pulsar RM data (Fig.~\ref{fig:han}) give a
significantly smaller pitch angle of -8\textdegree{} for the local
magnetic field \citep{han94,heiles96}. It is possible that the local
field forms a \emph{magnetic arm\/} located between two optical
arms, as in NGC~6946 (Sect.~4). Differences between the pitch angles
of the field and of the adjacent spiral arms of
10\textdegree{}--20\textdegree{} were also found in the spiral
galaxy M~51 \citep{patrikeev06}.

A model of the Galactic field based on 554 pulsar RM values
collected from several telescopes indicated reversals at several
Galactic radii, possibly between each spiral arm and the adjacent
interarm region \citep{han06}. The fields in the main inner arms
(Carina-Sagittarius, Scutum-Crux and Norma) run counterclockwise,
while the fields run clockwise in the interarm regions, in the solar
neighborhood and in the outer Perseus arm. On the other hand,
independent measurements of 150 pulsar RMs with the Parkes telescope
\citep{noutsos08} found safe evidence for only two reversals. The
first is located about 1~kpc inside the solar radius, between the
local field (clockwise field) and the Carina arm
(counter-clockwise). The second reversal occurs between the Carina
and Crux arms at 2--4~kpc distance from the sun towards the Galactic
center. The field in the Carina arm reverses from counter-clockwise
to clockwise beyond about 4~kpc distance from the sun
\citep{noutsos08}, contrary to \citep{han06}. However, sub-samples
of RM values above and below the Galactic plane in several regions
revealed different field reversals, which calls for caution with
interpreting the data.

To account for these reversals, a circular or axisymmetric field
with radial reversals \citep{vallee05} or a bisymmetric magnetic
spiral with a small pitch angle have been proposed
\citep{han94,han02,han06} (Fig.~\ref{fig:han}). However, none of the
models survived a statistical test \citep{men08}. The Milky Way's
field must have a complex structure which can be revealed only based
on much more data.

For example, all Galactic field models so far assumed a constant
pitch angle. However, experience from external galaxies shows that
the field's pitch angle is not constant along the spiral arm and
that the field may even slide away from the spiral arms, as observed
e.g. in M~51 \citep{patrikeev06}, but this cannot explain the
discrepancy between the Galactic and extragalactic observations.

The existence of large-scale reversals in the Milky Way is puzzling.
Very few such field reversals have been detected in external spiral
galaxies. Maps of Faraday rotation of the diffuse polarized
synchrotron emission are available for a couple of external spiral
galaxies. Though the spatial resolution with present-day radio
telescopes in external galaxies is lower, typically a few 100~pc,
large-scale field reversals in the RM maps of diffuse polarized
emission should be easily observable. Large-scale reversals were
found only in three galaxies. In the flocculent galaxy NGC~4414
\citep{soida02} and in the barred galaxy NGC~1097 \citep{beck05b}
the line of reversal runs at about constant azimuthal angle,
different from the reversals claimed for the Milky Way. A
bisymmetric field structure with a large pitch angle and two
azimuthal reversals on opposite sides of the disk possibly exists in
M~81 \citep{krause89b}. The disk fields of several galaxies can be
described by a mixture of dynamo modes, which may appears as a
radial reversal to an observer located within the disk
\citep{shukurov05}. However, no multiple reversals along radius,
like those in the Milky Way, were found so far in any external
galaxy. Either the Galactic reversals are not coherent over several
kpc, or they are restricted to a thin region near the Galactic
plane.

Some of the Galactic reversals may not be of Galactic extent, but
due to local field distortions or loops of the anisotropic turbulent
field. Pulsar RMs around a star formation complex indeed revealed a
field distortion which may mimic the reversal claimed to exist in
the direction of the Perseus arm \citep{mitra03,wielebinski05}.

Reversals on smaller scales are probably more frequent but also more
difficult to observe in external galaxies, because higher resolution
decreases the signal-to-noise ratio. Only in the barred galaxy
NGC~7479, where a jet serves as a bright polarized background,
several reversals on 1--2~kpc scale could be detected in the
foreground disk of the galaxy \citep{laine08}.

The discrepancy between Galactic and extragalactic data may also be
due to the different observational methods. RMs in external galaxies
are averages over the line of sight through the whole disk and halo
and over the large telescope beam, and they may miss field reversals
if these are restricted to a thin disk near the galaxy plane. The
results in the Milky Way are based on RMs of pulsars, which trace
the warm magneto-ionic medium near the plane along narrow lines of
sight.

Little is known about the vertical structure of the magnetic field
in the Milky Way. The vertical full equivalent thickness of the
thick radio disk of the Milky Way (up to 10~kpc radius) is
$\simeq2.2$~kpc near the Sun \citep{beuermann85} (scaled to a
distance to the Galactic center of 8~kpc), corresponding to an
exponential scale height of $1.6\pm0.2$~kpc which is very similar to
that in spiral galaxies (Sect.~6). The local Galactic field has only
a weak vertical component of $B_z\simeq0.2~\mu$G \citep{han94}.

Dynamo models predict the generation of quadrupole or dipole fields
where the toroidal component (traced by RMs) is symmetric or
antisymmetric with respect to the disk plane (Sect.~3). In the Milky
Way, RMs of extragalactic sources and of pulsars reveal no
large-scale reversal across the plane for Galactic longitudes
l=90\textdegree{}--270\textdegree{}. Thus the local field is part of
a large-scale symmetric (quadrupole) field structure parallel to the
Galactic plane. Towards the inner Galaxy
(l=270\textdegree{}--90\textdegree{}) the RM signs are opposite
above and below the plane (Fig.~\ref{fig:johnston}). This may
indicate a global antisymmetric (dipole) mode \citep{han97}. Sun et
al. \citep{sun08} confirmed from their recent analysis that the halo
field reverses its direction across the plane. However, the
existence of a superposition of a disk field with even parity and a
halo field with odd parity cannot be explained by classical dynamo
theory \citep{moss08}.

While the large-scale field is much more difficult to measure in the
Milky Way than in external galaxies, Galactic observations can trace
magnetic structures to much smaller scales \citep{reich06}. The
1.4~GHz all-sky polarization survey \citep{wolleben06} reveals a
wealth of structures on pc and sub-pc scales: filaments, canals,
lenses and rings. Their common property is to appear only in the
maps of polarized intensity, but not in total intensity. Some of
these are artifacts due to strong depolarization of background
emission in a foreground Faraday screen, called {\em Faraday
ghosts}\ and carry valuable information about the turbulent ISM in
the Faraday screen \citep{fletcher06}. Other features are associated
with real objects, like planetary nebulae \citep{ransom08} or the
photodissociation regions of molecular clouds \citep{wolleben04}.

\section{12. Summary and Outlook}

\begin{figure}[!t]
\includegraphics[bb = 48 40 561 388,width=1.0\columnwidth,clip=]{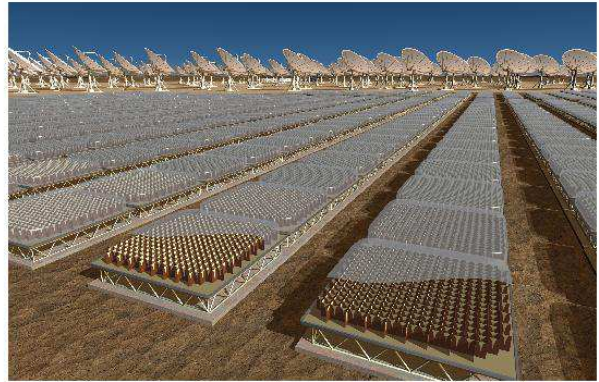}
\caption{SKA reference design: aperture array for low frequencies
and parabolic dishes for high frequencies (Copyright: SKA Programme
Development Office and XILOSTUDIOS).} \label{fig:ska}
\end{figure}

Radio synchrotron emission is a powerful tool to study the physics
of galaxy disks, halos and clusters. Ordered fields in galaxy disks
traced by polarized emission generally form spiral patterns, as
predicted by mean-field dynamo models. In galaxies with strong
non-axisymmetric gas flows around massive bars, the ordered fields
seem to follow the flow and can be used to map the velocity
components of the gas flow in the sky plane which are not accessible
to spectroscopic measurements.

The extent and spectral index of radio halos around galaxies and
galaxy clusters give information on the transport of cosmic-ray
electrons and their origin. Estimates of the electron lifetime allow
an estimate of their propagation speed. Fields compressed by galaxy
or cluster interactions or by ram pressure with the intergalactic
gas are observable via polarized emission. The shape of the galaxy
halos and their field structure allow to distinguish a global wind
driven by star formation in the disk from galactic fountain or
diffusion models. Faraday rotation measures trace regular fields and
can test dynamo models. Faraday rotation and depolarization are also
sensitive tools to detect ionized gas in galaxies, clusters and in
intergalactic space.

\begin{figure}[!t]
\includegraphics[bb = 48 40 561
732,width=0.9\columnwidth,clip=]{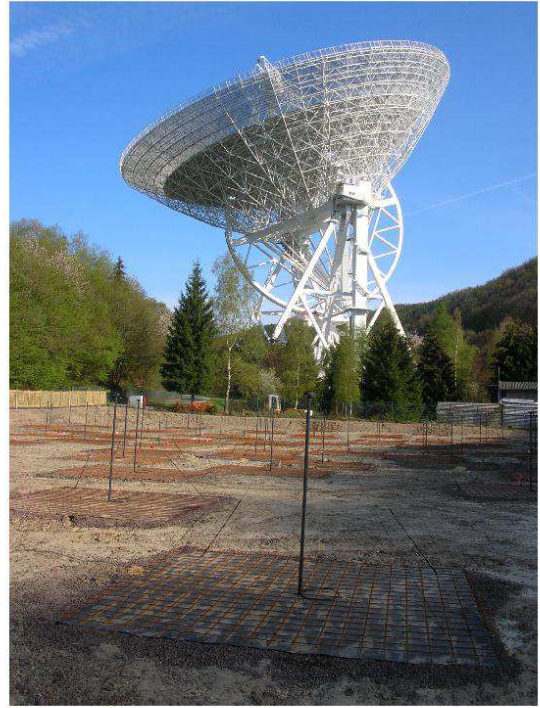}
\caption{The first international LOFAR station (front) with 96
low-band antennas (30--80~MHz) next to the Effelsberg 100~m
telescope (back) (Copyright: MPIfR Bonn).} \label{fig:lofar}
\end{figure}

Future radio telescopes will widen the range of observable magnetic
phenomena. High-resolution, deep observations at high frequencies
with the Extended Very Large Array (EVLA) and the planned Square
Kilometre Array (SKA) (Fig.~\ref{fig:ska}) are required to show
whether the ISM and halo fields are composed of many sheared loops
or are regular with dynamo-type patterns \citep{stepanov08}. The SKA
will allow to measure the Zeeman effect in much weaker magnetic
fields in the Milky Way and nearby galaxies. Forthcoming
low-frequency radio telescopes like the Low Frequency Array (LOFAR,
Fig.~\ref{fig:lofar}), Murchison Widefield Array (MWA), Long
Wavelength Array (LWA) and the low-frequency SKA will be suitable
instruments to search for extended synchrotron radiation at the
lowest possible levels in outer galaxy disks, halos and clusters,
and the transition to intergalactic space. At low frequencies we
will get access to the so far unexplored domain of weak magnetic
fields in galaxy halos \citep{beck08} and the population of cluster
halos with very steep radio spectra \citep{brunetti08,cassano08}.
The detection of radio emission from the intergalactic medium will
allow to probe the existence of magnetic fields in such rarified
regions, measure their intensity, and investigate their origin and
their relation to the structure formation in the early Universe.
``Cosmic magnetism'' is the title of Key Science Projects for LOFAR
and SKA \citep{beck07b}.

Multichannel spectro-polarimetric radio data allow to apply the
method of RM Synthesis \citep{brentjens05}. Faraday depolarization
can be reduced and features at different distances along the line of
sight can be separated. If the medium has a relatively simple
structure, for example a few emitting regions and Faraday screens,
{\em Faraday tomography}\ will become possible. This method will be
applied to all forthcoming polarization observations.

A reliable model for the global structure of the magnetic field of
the Milky Way needs a much higher number of pulsar and extragalactic
RM, hence much larger sensitivity. The SKA ``Magnetism'' Key Science
Project plans to observe an all-sky RM grid which should contain
about $10^4$ pulsar values with a mean spacing of $\simeq 30'$ and
obtain RMs from most of these \citep{gaensler04}. This survey will
be also be used to model the structure and strength of the magnetic
fields in intervening galaxies and clusters and in the intergalactic
medium \citep{beck04}. Deep polarization and RM observations will
shed light on the origin and evolution of cosmic magnetic fields.

\bibliographystyle{aipprocl} 

\begin{thebibliography}{9}

\bibitem{arshakian08}
T.~G.~Arshakian, R.~Beck, M.~Krause, and D.~Sokoloff, \emph{A\&A},
in press, arXiv:0810.3114 (2008).

\bibitem{balbus08}
S.~A.~Balbus, and C.~S.~Reynolds, \emph{ApJ} \textbf{681}, L65--L68
(2008).

\bibitem{bary87}
I.~Baryshnikova, A.~Shukurov, A.~Ruzmaikin, and D.~D.~Sokoloff,
\emph{A\&A} \textbf{177}, 27--41 (1987).

\bibitem{beck05a}
R.~Beck, in \emph{Cosmic Magnetic Fields}, eds. R.~Wielebinski and
R.~Beck, Springer, Berlin, 2005, pp.~41--68.

\bibitem{beck07a}
R.~Beck, \emph{A\&A} \textbf{470}, 539--556 (2007).

\bibitem{beck07b}
R.~Beck, \emph{Adv. Radio Sci.} \textbf{5}, 399--405 (2007).

\bibitem{beck08}
R.~Beck, \emph{Rev. Mex. AyA}, in press, arXiv:0804.4594 (2008).

\bibitem{beck96}
R.~Beck, A.~Brandenburg, D.~Moss, A.~Shukurov, and D.~Sokoloff,
\emph{Ann. Rev. A\&A} \textbf{34}, 155--206 (1996).

\bibitem{beck94a}
R.~Beck, C.~L.~Carilli, M.~A.~Holdaway, and U.~Klein, \emph{A\&A}
\textbf{292}, 409--424 (1994).

\bibitem{beck05b}
R.~Beck, A.~Fletcher, A.~Shukurov, {et~al.}, \emph{A\&A}
\textbf{444}, 739--765 (2005).

\bibitem{beck04}
R.~Beck, and B.~M.~Gaensler, in \emph{Science with the Square
Kilometer Array}, eds. C.~Carilli and S.~Rawlings, \emph{New Astr.
Rev.} \textbf{48}, Elsevier, Amsterdam, 2004, pp.~1289--1304.

\bibitem{beckkrause05}
R.~Beck, and M.~Krause, \emph{Astron. Nachr.} \textbf{326}, 414--427
(2005).

\bibitem{beck94b}
R.~Beck, A.~D.~Poezd, A.~Shukurov, and D.~D.~Sokoloff, \emph{A\&A}
\textbf{289}, 94--100 (1994).

\bibitem{beck03}
R.~Beck, A.~Shukurov, D.~Sokoloff, and R.~Wielebinski, \emph{A\&A}
\textbf{411}, 99--107 (2003).

\bibitem{bell03}
E.~F.~Bell, \emph{ApJ} \textbf{586}, 794--813 (2003).

\bibitem{berk71}
E.~M.~Berkhuijsen, \emph{A\&A} \textbf{14}, 359--386 (1971).

\bibitem{bernet08}
M.~L.~Bernet, F.~Miniati, S.~J.~Lilly, P.~P.~Kronberg, and
M.~Dessauges-Zavadsky, \emph{Nature} \textbf{454}, 302--304 (2008).

\bibitem{bertone06}
S.~Bertone, C.~Vogt, and T.~A.~En{\ss}lin, \emph{MNRAS}
\textbf{370}, 319--330 (2006).

\bibitem{beuermann85}
K.~Beuermann, G.~Kanbach, and E.~M.~Berkhuijsen, \emph{A\&A}
\textbf{153}, 17--34 (1985).

\bibitem{birk98}
G.~T.~Birk, H.~ Lesch, and T.~Neukirch, \emph{MNRAS} \textbf{296},
165--172 (1998).

\bibitem{brand93}
A.~Brandenburg, K.~J.~Donner, D.~Moss, {et~al.}, \emph{A\&A}
\textbf{271}, 36--50 (1993).

\bibitem{brand05}
A.~Brandenburg, and K.~Subramanian, \emph{Phys. Rep.} \textbf{417},
1--209 (2005).

\bibitem{brentjens05}
M.~A.~Brentjens, and A.~G.~de Bruyn, \emph{A\&A} \textbf{441},
1217--1228 (2005).

\bibitem{brown07}
J.~C.~Brown, M.~Haverkorn, B.~M.~Gaensler, {et~al.}, \emph{ApJ}
\textbf{663}, 258--266 (2007).

\bibitem{brunetti08}
G.~Brunetti, S.~Giacintucci, R.~Cassano, {et~al.}, \emph{Nature}
\textbf{455}, 944--947 (2008).

\bibitem{carilli02}
C.~L.~Carilli, and G.~B.~Taylor, \emph{Ann. Rev. A\&A} \textbf{40},
319--348 (2002).

\bibitem{cassano08}
R.~Cassano, G.~Brunetti, T.~Venturi, {et~al.}, \emph{A\&A}
\textbf{480}, 687--697 (2008).

\bibitem{chyzy04}
K.~T.~Chy{\.z}y, and R.~Beck, \emph{A\&A} \textbf{417}, 541--555
(2004).

\bibitem{chyzy07}
K.~T.~Chy{\.z}y, D.~J.~Bomans, M.~Krause, {et~al.}, \emph{A\&A}
\textbf{462}, 933--941 (2007).

\bibitem{chyzy08}
K.~T.~Chy{\.z}y, and R.~J.~Buta, \emph{ApJ} \textbf{677}, L17--L20
(2008).

\bibitem{chyzy06}
K.~T.~Chy{\.z}y, M.~Soida, D.~J.~Bomans, {et~al.}, \emph{A\&A}
\textbf{447}, 465--472 (2006).

\bibitem{cordes04}
J.~M.~Cordes, M.~Kramer, T.~J.~W.~Lazio, B.~W.~Stappers,
D.~C.~Backer, and S.~Johnston, in \emph{Science with the Square
Kilometer Array}, eds. C.~Carilli and S.~Rawlings, New Astronomy
Rev. \textbf{48}, Elsevier, Amsterdam, 2004, pp.~1413--1438.

\bibitem{crutcher99}
R.~M.~Crutcher, \emph{ApJ} \textbf{520}, 706--713 (1999).

\bibitem{crutcher07}
R.~M.~Crutcher, in \emph{Magnetic Fields in the Non-masing ISM},
eds. J.~M.~Chapman and W.~A.~Baan, Cambridge Univ. Press, Cambridge,
2007, pp.~47--54.

\bibitem{crutcher04}
R.~M.~Crutcher, D.~J.~Nutter, D.~Ward-Thompson, and J.~M.~Kirk,
\emph{ApJ} \textbf{600}, 279--285 (2004).

\bibitem{dahlem95}
M.~Dahlem, U.~Lisenfeld, and G.~Golla, \emph{ApJ} \textbf{444},
119--128 (1995).

\bibitem{avillez05}
M.~A.~de~Avillez, and D.~Breitschwerdt, \emph{A\&A} \textbf{436},
585--600 (2005).

\bibitem{vecchia08}
C.~Dalla Vecchia, and J.~Schaye, \emph{MNRAS} \textbf{387},
1431--1444 (2008).

\bibitem{dobbs08}
C.~L.~Dobbs, and D.~J.~Price, \emph{MNRAS} \textbf{383}, 497--512
(2008).

\bibitem{dumke98}
M.~Dumke, and M.~Krause, in \emph{The Local Bubble and Beyond}, eds.
D.~Breitschwerdt, M.~J.~Freyberg, and J.~Tr{\"u}mper, Springer,
Berlin, 1998, pp.~555--558 (1998).

\bibitem{dumke95}
M.~Dumke, M.~Krause, R.~Wielebinski, and U.~Klein, \emph{A\&A}
\textbf{302}, 691--703 (1995).

\bibitem{ehle98}
M.~Ehle, W.~Pietsch, R.~Beck, and U.~Klein, \emph{A\&A}
\textbf{329}, 39--54 (1998).

\bibitem{elstner92}
D.~Elstner, R.~Meinel, and R.~Beck, \emph{A\&A Suppl.} \textbf{94},
587--600 (1992).

\bibitem{ensslin98}
T.~A.~En{\ss}lin, P.~L.~Biermann, U.~Klein, and S.~Kohle \emph{A\&A}
\textbf{332}, 395--409 (1998).

\bibitem{ferriere00}
K.~Ferri\`ere, and D.~Schmitt, \emph{A\&A} \textbf{358}, 125--143
(2000).

\bibitem{fish03}
V.~L.~Fish, M.~J.~Reid, A.~L.~Argon, and K.~M.~Menten, \emph{ApJ}
\textbf{596}, 328--343 (2003).

\bibitem{fletcher04}
A.~Fletcher, E.~M.~Berkhuijsen, R.~Beck, and A.~Shukurov,
\emph{A\&A} \textbf{414}, 53--67 (2004).

\bibitem{fletcher06}
A.~Fletcher, and A.~Shukurov, \emph{MNRAS} \textbf{371}, L21--L25
(2006).

\bibitem{fomalont89}
E.~B.~Fomalont, K.~A.~Ebneter, W.~J.~M.~van~Breugel, and
R.~D.~Ekers, \emph{ApJ} \textbf{346}, L17--L20 (1989).

\bibitem{frick01b}
P.~Frick, R.~Beck, E.~M.~Berkhuijsen, I.~Patrikeev, \emph{MNRAS}
\textbf{327}, 1145--1157 (2001).

\bibitem{frick01a}
P.~Frick, R.~Stepanov, A.~Shukurov, and D.~Sokoloff, \emph{MNRAS}
\textbf{325}, 649--664 (2001).

\bibitem{gaensler04}
B.~M.~Gaensler, R.~Beck, and L.~Feretti, in \emph{Science with the
Square Kilometer Array}, eds. C.~Carilli and S.~Rawlings, \emph{New
Astr. Rev.} \textbf{48}, Elsevier, Amsterdam, 2004, pp.~1003--1012.

\bibitem{gaensler05}
B.~M.~Gaensler, M.~Haverkorn, L.~Staveley-Smith, {et~al.},
\emph{Science} \textbf{307}, 1610--1612 (2005).

\bibitem{giovannini91}
G.~Giovannini, L.~Feretti, and C.~Stanghellini, \emph{A\&A}
\textbf{252}, 528--537 (1991).

\bibitem{golla94}
G.~Golla, and E.~Hummel, \emph{A\&A} \textbf{284}, 777--792 (1994).

\bibitem{gomez02}
G.~C.~G\'omez, and D.~P.~Cox, \emph{ApJ} \textbf{580}, 235--252
(2002).

\bibitem{govoni04}
F.~Govoni, and L.~Feretti, \emph{Int. J. Mod. Phys. D} \textbf{13},
1549--1594 (2004).

\bibitem{govoni05}
F.~Govoni, M.~Murgia, L.~Feretti, {et al.}, \emph{A\&A}
\textbf{430}, L5--L8 (2005).

\bibitem{greaves00}
J.~S.~Greaves, W.~S.~Holland, T.~Jenness, and T.~G.~Hawarden,
\emph{Nature} \textbf{404}, 732--733 (2000).

\bibitem{gressel08}
O.~Gressel, D.~Elstner, U.~Ziegler, and G.~R\"udiger, \emph{A\&A}
\textbf{486}, L35--L38 (2008).

\bibitem{han98}
J.~L.~Han, R.~Beck, and E.~M.~Berkhuijsen, \emph{A\&A} \textbf{335},
1117--1123 (1998).

\bibitem{han97}
J.~L.~Han, R.~N.~Manchester, E.~M.~Berkhuijsen, and R.~Beck,
\emph{A\&A} \textbf{322}, 98--102 (1997).

\bibitem{han02}
J.~L.~Han, R.~N.~Manchester, A.~G.~Lyne, and G.~J.~Qiao, \emph{ApJ}
\textbf{570}, L17--L20 (2002).

\bibitem{han06}
J.~L.~Han, R.~N.~Manchester, A.~G.~Lyne, G.~J.~Qiao, and
W.~van~Straten, \emph{ApJ} \textbf{642}, 868--881 (2006).

\bibitem{han94}
J.~L.~Han, and G.~J.~Qiao, \emph{A\&A} \textbf{288}, 759--772
(1994).

\bibitem{han07}
J.~L.~Han, and J.~S.~Zhang, \emph{A\&A} \textbf{464}, 609--614
(2007).

\bibitem{hanasz04}
M.~Hanasz, G.~Kowal, K.~Otmianowska-Mazur, and H.~Lesch, \emph{ApJ}
\textbf{605}, L33--L36 (2004).

\bibitem{hanasz98}
M.~Hanasz, and H.~Lesch, \emph{A\&A} \textbf{332}, 77--87 (1998).

\bibitem{hanasz02}
M.~Hanasz, K.~Otmianowska-Mazur, and H.~Lesch, \emph{A\&A}
\textbf{386}, 347--358 (2002).

\bibitem{heckman90}
T.~M.~Heckman, L.~Armus, and G.~K.~Miley, \emph{ApJ Suppl.}
\textbf{74}, 833--868 (1990).

\bibitem{heesen08}
V.~Heesen, R.~Beck, M.~Krause, and R.-J.~Dettmar, \emph{A\&A}, in
press, arXiv:0812.0346 (2008).

\bibitem{heesen05}
V.~Heesen, M.~Krause, R.~Beck, and R.-J.~Dettmar, in \emph{The
Magnetized Plasma in Galaxy Evolution}, eds. K.~T.~Chyzy,
K.~Otmianowska-Mazur, M.~Soida, and R.-J.~Dettmar, Jagiellonian
University, Krak{ó}w, 2005, pp.~156--161.

\bibitem{heiles96}
C.~Heiles, in \emph{Polarimetry of the Interstellar Medium}, eds.
W.~G.~Roberge and D.~C.~B.~Whittet, ASP Conf. Ser. \textbf{97},
Astr. Soc. Pac., San Francisco, 1996, pp.~457--474.

\bibitem{heiles05}
C.~Heiles, and T.~H.~Troland, \emph{ApJ} \textbf{624}, 773--793
(2005).

\bibitem{heitsch04}
F.~Heitsch, E.~G.~Zweibel, A.~D.~Slyz, and J.~E.~G.~Devriendt,
\emph{ApJ} \textbf{603}, 165--179 (2004).

\bibitem{helou93}
G.~Helou, and M.~D.~Bicay, \emph{ApJ} \textbf{415}, 93--100 (1993).

\bibitem{hoernes98}
P.~Hoernes, E.~M.~Berkhuijsen, and C.~Xu, \emph{A\&A} \textbf{334},
57--70 (1998).

\bibitem{hughes06}
A.~Hughes, T.~Wong, R.~Ekers, {et~al.}, \emph{MNRAS} \textbf{370},
363--379 (2006).

\bibitem{hummel91a}
E.~Hummel, R.~Beck, and M.~Dahlem, \emph{A\&A} \textbf{248}, 23--29
(1991).

\bibitem{hummel91b}
E.~Hummel, R.~Beck, and R.-J.~Dettmar, \emph{A\&A Suppl.}
\textbf{87}, 309--317 (1991).

\bibitem{hummel90}
R.~Hummel, and R.-J.~Dettmar, \emph{A\&A} \textbf{236}, 33--46
(1990).

\bibitem{irwin99}
J.~A.~Irwin, J.~English, and B.~Sorathia, \emph{AJ} \textbf{117},
2102--2140 (1999).

\bibitem{johnston04}
M.~Johnston-Hollitt, C.~P.~Hollitt, and R.~D.~Ekers, in \emph{The
Magnetized Interstellar Medium}, eds. B.~Uyaniker, W.~Reich, and
R.~Wielebinski, Copernicus, Katlenburg, 2004, pp.~13--18.

\bibitem{klein88}
U.~Klein, R.~Wielebinski, and H.~W.~Morsi, \emph{A\&A} \textbf{190},
41--46 (1988).

\bibitem{kothes08}
R.~Kothes, T.~L.~Landecker, W.~Reich, S.~Safi-Harb, and
Z.~Arzoumanian, \emph{ApJ} \textbf{687}, 516--531 (2008).

\bibitem{krause90}
M.~Krause, in \emph{Galactic and Intergalactic Magnetic Fields},
eds. R.~Beck, R.~Wielebinski, and P.~P.~Kronberg, Kluwer, Dordrecht,
1990, pp.~187--196.

\bibitem{krause04}
M.~Krause, in \emph{The Magnetized Interstellar Medium}, eds.
B.~Uyaniker, W.~Reich, and R.~Wielebinski, Copernicus, Katlenburg,
2004, pp.~173--182.

\bibitem{krause07}
M.~Krause, \emph{Mem. Soc. Astr. Italiana} \textbf{78}, 314--316
(2007).

\bibitem{krause08}
M.~Krause, \emph{Rev. Mex. AyA}, in press, arXiv:0806.2060 (2008).

\bibitem{krause89b}
M.~Krause, R.~Beck, and E.~Hummel, \emph{A\&A} \textbf{217}, 17--30
(1989).

\bibitem{krause89a}
M.~Krause, E.~Hummel, and R.~Beck, \emph{A\&A} \textbf{217}, 4--16
(1989).

\bibitem{krause06}
M.~Krause, R.~Wielebinski, and M.~Dumke, \emph{A\&A} \textbf{448},
133--142 (2006).

\bibitem{kronberg94}
P.~P.~Kronberg, \emph{Rep. Prog. Phys.} \textbf{57}, 325--382
(1994).

\bibitem{kronberg06}
P.~P.~Kronberg, \emph{Astr. Nachr.} \textbf{327}, 517--522 (2006).

\bibitem{kronberg08}
P.~P.~Kronberg, M.~L.~Bernet, F.~Miniati, S.~J.~Lilly, M.~B.~Short,
and D.~M.~Higdon, \emph{ApJ} \textbf{676}, 70--79 (2008).

\bibitem{kronberg07}
P.~P.~Kronberg, R.~Kothes, C.~J.~Salter, and P.~Perillat, \emph{ApJ}
\textbf{659}, 267--274 (2007).

\bibitem{kronberg99}
P.~P.~Kronberg, H.~Lesch, and U.~Hopp, \emph{ApJ} \textbf{511},
56--64 (1999).

\bibitem{kronberg92}
P.~P.~Kronberg, J.~J.~Perry, and E.~L.~H.~Zukowski, \emph{ApJ}
\textbf{387}, 528--535 (1992).

\bibitem{laine08}
S.~Laine, and R.~Beck, \emph{ApJ} \textbf{673}, 128--142 (2008).

\bibitem{larosa05}
T.~N.~LaRosa, C.~L.~Brogan, S.~N.~Shore, T.~J.~Lazio, N.~E.~Kassim,
and M.~E.~Nord, \emph{ApJ} \textbf{626}, L23--L27 (2005).

\bibitem{lisenfeld96a}
U.~Lisenfeld, H.~J.~V\"olk, and C.~Xu, \emph{A\&A} \textbf{306},
677--690 (1996).

\bibitem{lisenfeld96b}
U.~Lisenfeld, H.~J.~V\"olk, and C.~Xu, \emph{A\&A} \textbf{314},
745--753 (1996).

\bibitem{men08}
H.~Men, K.~Ferri{\`e}re, and J.~L.~Han, \emph{A\&A} \textbf{486},
819--828 (2008).

\bibitem{mitra03}
D.~Mitra, R.~Wielebinski, M.~Kramer, and A.~Jessner, \emph{A\&A}
\textbf{398}, 993--1005 (2003).

\bibitem{moss99}
D.~Moss, A.~Shukurov, and D.~Sokoloff, \emph{A\&A} \textbf{343},
120--131 (1999).

\bibitem{moss08}
D.~Moss, and D.~Sokoloff, \emph{A\&A} \textbf{487}, 197--203 (2008).

\bibitem{nieten06}
Ch.~Nieten, N.~Neininger, M.~Gu{\'e}lin, {et~al.}, \emph{A\&A}
\textbf{453}, 459--475 (2006).

\bibitem{niklas97}
S.~Niklas, and R.~Beck, \emph{A\&A} \textbf{320}, 54--64 (1997).

\bibitem{noutsos08}
A.~Noutsos, S.~Johnston, M.~Kramer, and A.~Karastergiou,
\emph{MNRAS} \textbf{386}, 1881--1896 (2008).

\bibitem{novak05}
G.~Novak, in \emph{Magnetic Fields in the Universe}, eds.
E.~M.~de~Gouveia Dal Pino et al., AIP Conf. Proc. \textbf{784},
Melville, 2005, pp.~329--339.

\bibitem{parker92}
E.~N.~Parker, \emph{ApJ} \textbf{401}, 137--145 (1992).

\bibitem{patrikeev06}
I.~Patrikeev, A.~Fletcher, R.~Stepanov, {et~al.}, \emph{A\&A}
\textbf{458}, 441--452 (2006).

\bibitem{pietsch00}
W.~Pietsch, A.~Vogler, U.~Klein, and H.~Zinnecker, \emph{A\&A}
\textbf{360}, 24--48 (2000).

\bibitem{pierini03}
D.~Pierini, C.~C.~Popescu, R.~J.~Tuffs, and H.~J.~V\"olk,
\emph{A\&A} \textbf{409}, 907--916 (2003).

\bibitem{pohl90}
M.~Pohl, and R.~Schlickeiser, \emph{A\&A} \textbf{234}, 147--155
(1990).

\bibitem{price08}
D.~J.~Price, and M.~R.~Bate, \emph{MNRAS} \textbf{385}, 1820--1834
(2008).

\bibitem{rand94}
R.~J.~Rand, and A.~G.~Lyne, \emph{MNRAS} \textbf{268}, 497--505
(1994).

\bibitem{ransom08}
R.~R.~Ransom, B.~Uyaniker, R.~Kothes, and T.~L.~Landecker,
\emph{ApJ} \textbf{684}, 1009--1017 (2008).

\bibitem{rees05}
M.~J.~Rees, in \emph{Cosmic Magnetic Fields}, eds. R.~Wielebinski
and R.~Beck, Springer, Berlin, 2005, pp.~1--8.

\bibitem{reich94}
W.~Reich, in \emph{The Nuclei of Normal Galaxies}, eds. R.~Genzel
and A.~I.~Harris, Kluwer, Dordrecht, 1994, pp.~55--62.

\bibitem{reich06}
W.~Reich, in \emph{Cosmic Polarization}, ed. R.~Fabbri, Research
Signpost, Kerala, 2006, pp.~91--130.

\bibitem{reuter92}
H.-P.~Reuter, U.~Klein, H.~Lesch, R.~Wielebinski, and
P.~P.~Kronberg, \emph{A\&A} \textbf{256}, 10--18 (1992).

\bibitem{reuter94}
H.-P.~Reuter, U.~Klein, H.~Lesch, R.~Wielebinski, and
P.~P.~Kronberg, \emph{A\&A} \textbf{282}, 724--730 (1994).

\bibitem{robishaw08}
T.~Robishaw, E.~Quataert, and C.~Heiles, \emph{ApJ} \textbf{680},
981--998 (2008).

\bibitem{roussel03}
H.~Roussel, G.~Helou, R.~Beck, {et~al.}, \emph{ApJ} \textbf{593},
733--759 (2003).

\bibitem{scarrott87}
S.~M.~Scarrott, D.~Ward-Thompson, and R.~F.~Warren-Smith,
\emph{MNRAS} \textbf{224}, 299--305 (1987).

\bibitem{seymour08}
N.~Seymour, T.~Dwelly, D.~Moss, {et~al.}, \emph{MNRAS} \textbf{386},
1695--1708 (2008).

\bibitem{shukurov05}
A.~Shukurov, in \emph{Cosmic Magnetic Fields}, eds. R.~Wielebinski
and R.~Beck, Springer, Berlin, 2005, pp.~113--135.

\bibitem{shukurov06}
A.~Shukurov, D.~Sokoloff, K.~Subramanian, and A.~Brandenburg,
\emph{A\&A} \textbf{448}, L33--L36 (2006).

\bibitem{soida02}
M.~Soida, R.~Beck, M.~Urbanik, M., and J.~Braine, \emph{A\&A}
\textbf{394}, 47--57 (2002).

\bibitem{sokoloff98} D.~D.~Sokoloff, A.~A.~Bykov, A.~Shukurov, E.~M.~Berkhuijsen,
R.~Beck, and A.~D.~Poezd, \emph{MNRAS} \textbf{299}, 189--206 (1998)
and Erratum in \emph{MNRAS} \textbf{303}, 207--208 (1999).

\bibitem{stepanov08}
R.~Stepanov, T.~G.~Arshakian, R.~Beck, P.~Frick, and M.~Krause,
\emph{A\&A} \textbf{480}, 45--59 (2008).

\bibitem{strong00}
A.~W.~Strong, I.~V.~Moskalenko, and O.~Reimer, \emph{ApJ}
\textbf{537}, 763--784 (2000).

\bibitem{subra98}
K.~Subramanian, \emph{MNRAS} \textbf{294}, 718--728 (1998).

\bibitem{subra06}
K.~Subramanian, A.~Shukurov, and N.~E.~L.~Haugen, \emph{MNRAS}
\textbf{366}, 1437--1454 (2006).

\bibitem{sun08}
X.~H.~Sun, W.~Reich, A.~Waelkens, and T.~A.~En{\ss}lin, \emph{A\&A}
\textbf{477}, 573--592 (2008).

\bibitem{taba07}
F.~Tabatabaei, R.~Beck, M.~Krause, et al., \emph{A\&A} \textbf{466},
509--519 (2007).

\bibitem{taba08}
F.~Tabatabaei, M.~Krause, A.~Fletcher, and R.~Beck, \emph{A\&A}
\textbf{490}, 1005--1017 (2008).

\bibitem{taylor93}
G.~B.~Taylor, and R.~A.~Perley, \emph{ApJ} \textbf{416}, 554--562
(1993).

\bibitem{thompson06}
T.~A.~Thompson, E.~Quataert, E.~Waxman, N.~Murray, and C.~L.~Martin,
\emph{ApJ} \textbf{645}, 186--198 (2006).

\bibitem{tuell06}
R.~T{\"u}llmann, D.~Breitschwerdt, J.~Rossa, W.~Pietsch, and
R.-J.~Dettmar, A\&A \textbf{457}, 779--785 (2006).

\bibitem{tuell00}
R.~T{\"u}llmann, R.-J.~Dettmar, M.~Soida, M.~Urbanik, and J.~Rossa,
\emph{A\&A} \textbf{364}, L36--L41 (2000).

\bibitem{vallee02}
J.~P.~Vall{\'e}e, \emph{ApJ} \textbf{566}, 261--266 (2002).

\bibitem{vallee05}
J.~P.~Vall{\'e}e, \emph{ApJ} \textbf{619}, 297--305 (2005).

\bibitem{vallee08}
J.~P.~Vall{\'e}e, \emph{ApJ} \textbf{681}, 303--310 (2008).

\bibitem{vazquez05}
E.~V{\'a}zquez-Semadeni, J.~Kim, and J.~Ballesteros-Paredes,
\emph{ApJ} \textbf{630}, L49--L52 (2005).

\bibitem{voelk89}
H.~J.~V\"olk, \emph{A\&A} \textbf{218}, 67--70 (1989).

\bibitem{vogler05}
A.~Vogler, S.~C.~Madden, R.~Beck, {et~al.}, \emph{A\&A}
\textbf{441}, 491--511 (2005).

\bibitem{vogt03}
C.~Vogt, and T.~A.~En{\ss}lin, \emph{A\&A} \textbf{412}, 373--385
(2003).

\bibitem{vogt05}
C.~Vogt, and T.~A.~En{\ss}lin, \emph{A\&A} \textbf{434}, 67--76
(2005).

\bibitem{vollmer07}
B.~Vollmer, M.~Soida, R.~Beck, {et~al.}, \emph{A\&A} \textbf{464},
L37--L40 (2007).

\bibitem{walsh02}
W.~Walsh, R.~Beck, G.~Thuma, {et~al.}, \emph{A\&A} \textbf{388},
7--28 (2002).

\bibitem{wez07}
M.~We{\.z}gowiec, M.~Urbanik, B.~Vollmer, {et~al.}, \emph{A\&A}
\textbf{471}, 93--102 (2007).

\bibitem{widrow02}
L.~M.~Widrow, \emph{Rev. Mod. Phys.} \textbf{74}, 775--823 (2002).

\bibitem{wielebinski05}
R.~Wielebinski, in \emph{Cosmic Magnetic Fields}, eds.
R.~Wielebinski and R.~Beck, Springer, Berlin, 2005, pp.~89--112.

\bibitem{wolfe08}
A.~M.~Wolfe, R.~A.~Jorgenson, T.~Robishaw, C.~Heiles, and
J.~X.~Prochaska, \emph{Nature} \textbf{455}, 638--640 (2008).

\bibitem{wolleben06}
M.~Wolleben, T.~L.~Landecker, W.~Reich, and R.~Wielebinski,
\emph{A\&A} \textbf{448}, 411--424 (2006).

\bibitem{wolleben04}
M.~Wolleben, and W.~Reich, \emph{A\&A} \textbf{427}, 537--548
(2004).

\bibitem{yusef96}
F.~Yusef-Zadeh, D.~A.~Roberts, W.~M.~Goss, D.~A.~Frail, and
A.~J.~Green, \emph{ApJ} \textbf{466}, L25--L29 (1996).

\bibitem{zimmer97}
F.~Zimmer, H.~Lesch, and G.~T.~Birk, \emph{A\&A} \textbf{320},
746--756 (1997).

\end{thebibliography}

\end{document}